\documentclass[12pt,a4paper]{article}
\usepackage[latin1]{inputenc}
\usepackage{amsmath}
\usepackage{amsfonts}
\usepackage{amssymb,graphicx,cite}
\newcommand{\be}{\begin{equation}}
\newcommand{\ee}{\end{equation}}

\begin{document}

\begin{center}
{\Large\bf Experimental indication on chiral symmetry restoration in meson spectrum}
\end{center}

\begin{center}
{\large S.S.~Afonin} \\
Departament d'ECM,
Universitat de Barcelona, \\
Barcelona 08028, Spain\\
e-mail: afonin@ecm.ub.es
\end{center}

\begin{abstract}
The spectroscopic predictions of the Ademollo-Veneziano-Weinberg dual model
are critically tested in view of the modern experimental data.
The predicted equidistance of masses squared for chiral partners
is shown to be violated high in energies, instead one observes
an approximate degeneracy of these quantities. This phenomenon
can be interpreted as the
restoration of Wigner-Weyl realization of chiral symmetry for
highly excited states. The scale of complete restoration is
expected to be 2.5~GeV. A multispin-parity cluster structure of
meson spectrum is revealed.
\end{abstract}

\noindent
Pacs: 12.90.+b, 12.38.-t,12.39.Mk \\
Keywords: parity doubling, chiral symmetry restoration

\section{Introduction}

The problem of parity doubling in the spectrum of high "radial"
excitations of light hadrons has been actively discussed
recently~\cite{golt3,jaffenew,G5,jaffe,we,sh,in,swan,golt2}. One
often relates this phenomenon to the Chiral Symmetry Restoration
(CSR) at high energies (see, e.g.,~\cite{kirch,G3,AV,pi}) which is
spontaneously broken at low energies. Namely, the typical scale of
Chiral Symmetry Breaking (CSB) is
$\Lambda_{\text{CSB}}\approx1$~GeV. Below this scale the chiral
symmetry is known to be realized nonlinearly (the Nambu-Goldstone
realization). But above this scale the linear (Wigner-Weyl)
realization of chiral symmetry is expected to be restored. Since
the chiral symmetry of QCD Lagrangian is almost exact in the light
quark sector, an experimental signal for CSR should be approximate
degeneracy of masses of chiral partners among light mesons above
1~GeV. Thus, investigations of parity doubling (both experimental
and theoretical) are expected to shed some light on the phenomenon
of CSB in QCD.

CSR in meson spectrum was predicted by the QCD sum
rules~\cite{we,sh,we2,bean} and some effective
models~\cite{weandG}. On the other hand, the possibility of such
phenomenon was criticized in~\cite{jaffenew,jaffe} (within the sum
rules in~\cite{golt3,golt2}). In particular, there exist other
explanations for parity doubling (for a review see~\cite{review},
in the baryon sector see~\cite{jaffenew,in}). The experimental
status of parity doubling is still ambiguous. While in the baryon
sector the situation is more or less certain (see,
e.g.,~\cite{in}), in the meson sector the data are rather scarce
to see the effect distinctly. However, some time ago the analysis
of the experimental data of Crystal Barrel Collaboration was
published~\cite{anietall} where many new meson resonances were
revealed. The obtained results were systematized
in~\cite{ani,ani2}. The comprehensive review is contained in~\cite{bugg}.
The experimental spectrum turned out to be
very similar to the spectrum typically given by effective boson
string theories.

In this Letter we show that the obtained
in~\cite{anietall,ani,ani2} data are consistent with CSR. As an auxiliary
tool we will exploit the Ademollo-Veneziano-Weinberg dual model~\cite{gLS}.

The paper is organized as follows. In Section~2 we argue that the
available experimental data favour CSR at high energies.
Section~3 is devoted to some discussions.
%an intuitive interpretation of
%multispin-parity clustering revealed in the spectrum.
The conclusions are summarized in final Section~4.

\section{Phenomenological analysis}

Let us discuss first the relation between CSR and parity doubling.
In the relativistic theories one always deals with (masses)$^2$ which
mark different representations of Lorentz group. Consequently, only
these quantities appear in all theoretical results. If chiral
symmetry is restored parity doublets fall into multiplets of the
chiral group with equal (mass)$^2$ values~\cite{gh}. Therefore,
{\it only approximate degeneracy of (masses)$^2$ of chiral partners
signalizes CSR in particle spectrum}. Thus, parity doubling is
necessary but not sufficient condition for CSR. A practical
consequence is that conclusions about CSR cannot be inferred from
analysis of masses (as some authors do), one should analyze
(masses)$^2$. The reason is trivial: If the (masses)$^2$ of
chiral partners grow, say, as $M_{\pm}^2(n)\sim n^{\alpha}$ ($\alpha>0$)
and the corresponding differences behave as
$M_+^2(n)-M_-^2(n)\sim n^{\beta}$ then at $\beta<\frac{\alpha}{2}$
one always has the parity doubling high enough in energy in the sense
$M_+(n)-M_-(n)\xrightarrow[n\rightarrow\infty]{}0$,
while the {\it genuine} CSR
obviously requires $\beta<0$. At $0\leqslant\beta<\frac{\alpha}{2}$
one deals with the {\it effective} CSR only.

Keeping in mind this distinction, let us proceed now to the
experimental evidence for parity doubling in the sector of light
mesons. Typically the evidences were based on some direct observations
showing that opposite parity states cluster nearby in mass.
However, it is often not clear which states are to be compared in
channels with many states. This was a reason why such a naive
treatment of experimental data was criticized in recent
work~\cite{jaffenew}, where a quite involved statistical method
was proposed measuring the correlations between
negative and positive parity states. This method was applied then
to baryons and a significant signal for parity doubling in the
non-strange baryons was revealed. From this work, however, we
cannot judge whether there is a signal for CSR as only
correlations between masses (not masses squared) were measured. In
addition, in the baryon sector the parity doubling can be
explained by restoration (or dynamical suppression) of axial
symmetry only.

We would like to propose a much simpler theoretical construction
for checking parity doubling and CSR in the light meson sector. In
order to see which states are chiral partners one can use some
phenomenologically successful model describing the meson spectrum.
Comparing the predictions of this model with the experimental
data, it may be possible to assess the parity
doubling. In our opinion, the best candidate for the light meson
sector is the dual resonance model proposed by Lovelace and
Shapiro in~\cite{LS} and generalized by Ademollo, Veneciano
and Weinberg (AVW) in~\cite{gLS}. The spectrum of this model is depicted
in Fig.~1. A characteristic feature of this spectrum is that
(masses)$^2$ of chiral partners always have a constant shift
(equal to $m_{\rho}^2$), i.e. in this case we have an {\it
effective parity doubling} high in spectrum because
\begin{equation}
M_+(n)-M_-(n)=\frac{m_{\rho}^2}{M_+(n)+M_-(n)}\longrightarrow 0,
\quad n\gg1.
\end{equation}
As discussed above, such an effective mass degeneracy high in spectrum is not a signal
for CSR.

\begin{center}
\begin{figure*}
\label{Fig}
\vspace{-5cm}
\hspace{-4cm}
\includegraphics[scale=1]{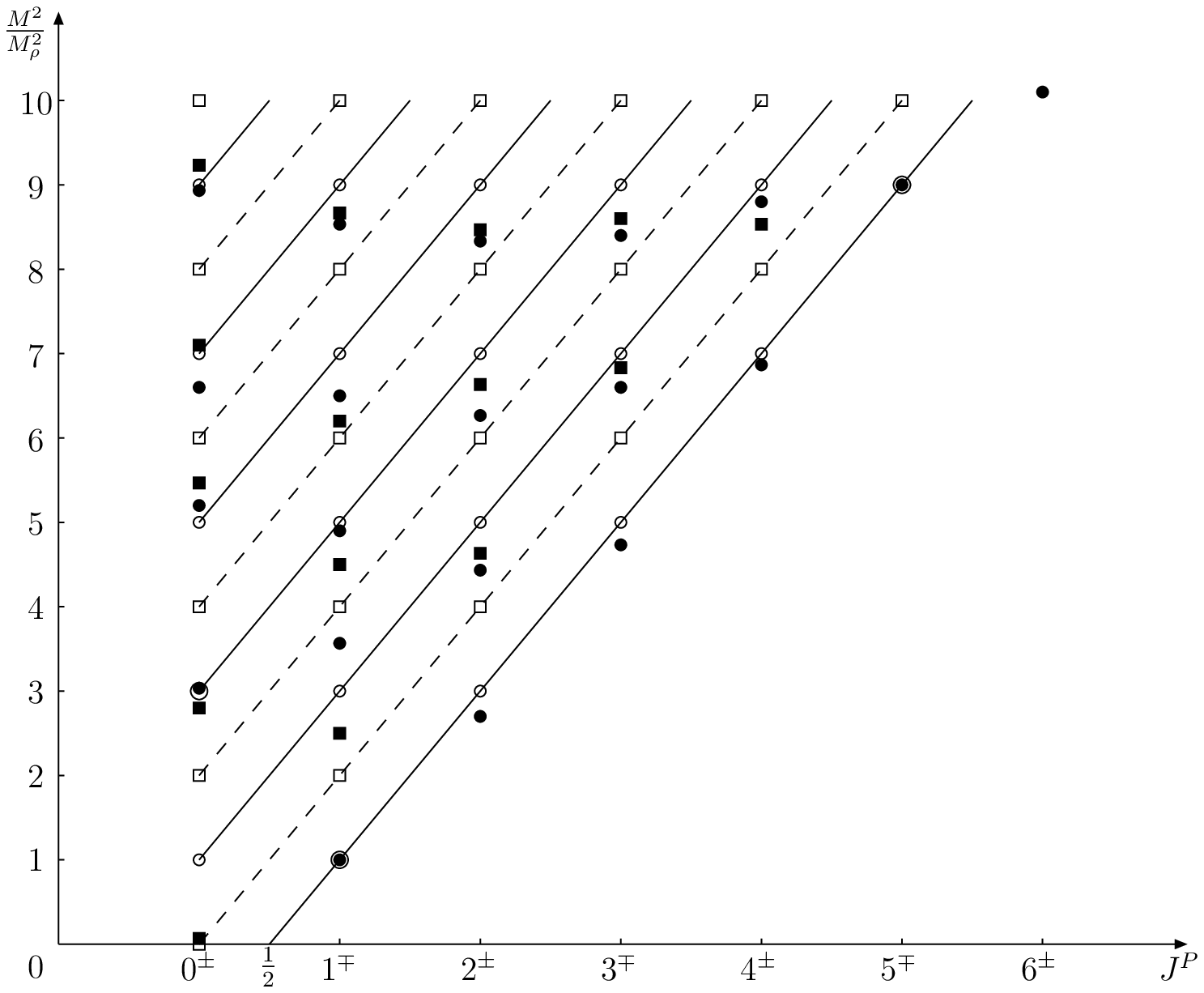}
\vspace{-13cm}
\caption{The Regge trajectories predicted by the AVW spectrum~\cite{gLS}
in units of $M_{\rho(770)}^2$:
$\rho$-meson trajectory with daughters (solid lines) and $\pi$-meson trajectory with daughters
(dashed lines). The predicted states are denoted by open circles and squares correspondingly.
The filled circles and squares represent the corresponding averaged experimental values.
The following experimental $J^{P}$ states were used~\cite{pdg} (the states discovered in
analysis~\cite{anietall,ani2} are marked by star).
$0^+$($f_0$-mesons): $?$, $1350\pm150$, $1770\pm12(*)$, $1992\pm16$, $2320\pm30(*)$;
$0^-$($\pi$-mesons): $140$, $1300\pm100$, $1812\pm14$, $2070\pm35(*)$, $2360\pm30(*)$;
$1^-$($\rho$-mesons): $775.8\pm0.5$, $1465\pm25$, $1720\pm20$, $1980\pm30(*)$, $2265\pm40(*)$;
$1^+$($a_1$-mesons): $1230\pm40$, $1647\pm22$, $1930^{+30}_{-70}(*)$, $2270^{+55}_{-40}(*)$;
$2^+$($f_2$-mesons): $1275\pm1$, $1638\pm6$, $1945\pm13$, $2240\pm30(*)$;
$2^-$($\pi_2$-mesons): $1672\pm3$, $2005\pm15(*)$, $2245\pm60(*)$;
$3^-$($\rho_3$-mesons): $1686\pm4$, $1980\pm15(*)$, $2250\pm?$;
$3^+$($a_3$-mesons): $2030\pm20(*)$, $2275\pm35(*)$;
$4^+$($f_4$-mesons): $2034\pm11$, $2300\pm?$;
$4^-$($\pi_4$-mesons): $2250\pm15(*)$;
$5^-$($\rho_5$-mesons): $2330\pm35$;
$6^+$($f_6$-mesons): $2465\pm50$.}
\end{figure*}
\end{center}
\vspace{-1.14cm}

We analyzed all available spectroscopic data
and depicted it in the same plot. After excluding all mesons
with a large admixture of strange quark (which is seen from the
corresponding decay channels) each experimentally detected state
can be confronted with the one predicted by the model. We have
found only a few exceptions: extra-states $\rho(1900)$, $\rho(2150)$,
$f_2(1565)$, $f_2(1810)$ and missing ground scalar meson.
The existence of additional states in $(I,J^{PC})=(I,1^{--}),\,(I,2^{++}),...$
channels is well known. In terms of the non-relativistic language their
appearance is due to the fact that these states can be created
by different orbital momenta~\cite{ani}, i.e. there exist not only
S-wave vector and P-wave tensor mesons but also D-wave vector and
F-wave tensor states. In the relativistic theories this seems to be related
with the existence of two interpolating currents for the corresponding
states~\cite{G3}.
%$\rho(1900)$ is,
%however, a very suspicious in any way: Its decay width is by order
%of magnitude less than the typical widths of light mesons with
%comparable masses, hence, it is really doubtful that this state
%may be considered on the same footing with others. $\rho(2150)$ is
%superfluous for the model under consideration and, thus, it might
%be considered as the first candidate for the vector meson of
%"second kind" in the spirit of~\cite{G3}.
The both candidates
for the lightest scalar state, $f_0(600)$ and $f_0(980)$ mesons,
seem not to be genuine quark-antiquark $SU_f(2)$ states (see note
on scalar mesons in Particle Data Group~\cite{pdg}). Thus, we
refrain from placing them on the trajectory.

As pointed out above, if the experimental states in Fig.~1
followed the prescribed positions we would have only effective
parity doubling. The situation in the real world is, however,
different: While the states with the quantum numbers of
$\rho$-meson trajectory and its daughters tend to occupy the
positions prescribed by the model, their chiral partners "glue"
them the stronger the higher in energy the excitations are
located. Other property of the experimental spectrum is that the
states lying on the principal $\rho$-meson Regge trajectory
have no chiral partners. These two features are general for
channels with any spin and, consequently, they hardly can be
accidental.

Effective parity doubling predicted by the model means that all
states are equally influenced by CSB. In this case we observe
parity doubling high in spectrum only because chirally
non-invariant contribution to the masses remains constant while
masses are growing, i.e. CSB effects become unimportant high in
energy. Such a behavior was discussed  within some approaches
based on QCD sum rules~\cite{golt3,garda,afon}. Experimental
spectrum depicted in Fig.~1 favours, however, another scenario:
Chirally non-invariant contribution gradually decreases so that
high in spectrum the (masses)$^2$ of chiral partners practically
coincide within experimental accuracy. In other words, {\it
experimental data favour a genuine CSR high in energy}.

A prominent feature of experimental spectrum, which we
like to emphasize, is the existence of well-pronounced
multispin-parity clusters of states, very similar to those observed
in the baryon spectrum~\cite{hol} (for a review see~\cite{in}).
Namely, four such clusters are
distinctively seen. The corresponding clustering occurs at $1350\pm120$,
$1720\pm90$, $2000\pm70$, and $2300\pm60$ MeV with approximately
the same mass gap between clusters,
$m^2_{\text{gap}}\approx(1080\,\text{MeV})^2\approx2m_{\rho}^2$.
The mass splitting within clusters is gradually decreasing high in
energy. The clustering of mesons into narrow mass ranges was observed
in a bit different aspect in~\cite{bugg}.

\section{Discussions}

Let us try to figure out (at least intuitively) the origin of
clusters in meson spectrum. To do this it is convenient to "switch
off" two space dimensions and for a while do not bother about some
nuances concerning the dynamics and the classification of states on
parity. Then the spin degrees of freedom will disappear and
%each
%cluster will consist of only two states with opposite parity.
the states in Fig.~1 will "collapse" to axis $0^{\pm}$. The
AVW spectrum in this case can be written in the form
$m^2(n)=m_{\rho}^2n$, with $n=0,1,2,\dots$, where the states alternate
in parity. It resembles the asymptotics of dim2 QCD spectrum in
the large-$N_c$ limit~\cite{dim2}. This spectrum is completely
determined by the confinement forces (in dim2 there is no spontaneous
CSB but the massless pion exists). If we now "switch on" two
additional space dimensions, each state then becomes
$1+(n+\delta_{1P})/2$ times degenerate (here $P=\pm1$ is parity
and the Kronecker symbol was used) giving rise to a cluster of
states. The permitted values for masses of excited states are
still determined by confinement, but these values can be now
obtained not only by exciting the ground state "radially" but also
by exciting "orbitally". The spin interactions generate then a
mass splitting within a cluster: Different spins have rather
different "sensitivity" to CSB which occurs in dim4 QCD. Namely,
the lower the spin is, the more sensitive it is to CSB (loosely
speaking, as spin is a manifestation of quantum "inner motion",
the faster this motion is, the less sensitive to the structure of
vacuum it is). Fig.~1 is in agreement with such a intuitive
picture: The masses of states with $J=0,1$ typically reveal the
largest deviations from the averaged mass within a cluster.
Nevertheless, the splitting is less than the mass gap between the
clusters which gives us a possibility to observe them. Another
mentioned property of clusters in Fig.~1 is that the higher are
the excitations the more pronounced cluster they form. This
signals that the CSB effects gradually disappear high in energy.
As a result we have $M^2(n,J)\sim (n+J)$ for large enough $n$ and
$J$. Such a behavior was predicted by the relativistic Nambu-Gotto
open string (see, e.g.,~\cite{zw}) and by some effective string
models (see, e.g.,~\cite{bak}). In addition, as was recently
reported in~\cite{kar}, it can be given by AdS/QCD for vector and
axial-vector mesons. However, these clusters are only {\it
multispin} ones. What actually happens in reality (at least in the
considered channels) is the {\it multispin-parity} clustering,
i.e. $M^2(P,n,J)=M^2(n+J)$. In terms of the linear ansatz it means
that at large $n$ and $J$ one has $M^2(P,n,J)=a(n+J)+b$ with
universal slope $a$ and intercept $b$. It is the degeneracy of
P-parity and/or chiral partners (they do not always coincide)
which indicates on CSR.
The AVW amplitude predicts
the universal slope $a=2m_{\rho}^2$ which is fulfilled
experimentally (this result was also derived within the QCD sum rules
in~\cite{we2,afon}). The universal intercept is very close to the
value $b=m_{\rho}^2$. As was shown in~\cite{we2} such an intercept
corresponds to a chirally-symmetric linear spectrum. This fact is
another independent indication on the {\it genuine CSR high in
meson spectrum}. However, CSR cannot completely describe the
multispin-parity clustering since it does not predict the degeneracy
of chiral multiplets with different spins. Some other symmetry is
responsible for this phenomenon.
Due to the Lorentz nature of spin
%The number of states in a given cluster
%is determined by value of energy where it occurs, i.e. by QCD
%dynamics (the higher energy is, the more ways exist to excite
%orbitally), and by number of space dimensions (the more dimensions
%are, the more orientations for spin are possible). Thus,
%dimensionality of space is of importance and, hence,
the phenomenon might be explained by an appropriate grouping the
mesons into some irreducible representations of Lorentz group (an
example in the baryon sector is given in~\cite{ki}).

A partial CSR is clearly observed at 1.7~GeV (position of the
second cluster) and then CSR is rapidly progressing.
If the tendency towards CSR seen in~Fig.~1 persists higher in
spectrum then at $\Lambda_{\text{CSR}}\approx2.5$~GeV (the position of
the fifth hypothetical cluster) we should observe a complete CSR
within the experimental errors. The same value for
$\Lambda_{\text{CSR}}$ was obtained in framework of rather
orthogonal approaches in~\cite{swan,kalashnikova} (in the latter paper
it was done for heavy-light quarkonia, but $\Lambda_{\text{CSR}}$
in that sector is expected to be the same). The physics above
$\Lambda_{\text{CSR}}$ is indistinguishable from the physics of
perturbative QCD continuum. The scale $\Lambda_{\text{CSR}}$ marks
the transition from intermediate energy to high energy like the
scale $\Lambda_{\text{CSB}}$ does from low to intermediate one.
Thus, if experimentally confirmed, the scale
$\Lambda_{\text{CSR}}$ is of a great importance because it marks
the upper bound of resonance physics in the light quark sector of
QCD. Theoretically it should be then considered as the third
important scale in QCD. A practical consequence of the existence
of this scale is that the most selfconsistent matching of finite
energy sum rules and effective field models with QCD continuum can
be achieved only at $\Lambda_{\text{CSR}}$.

Why the AVW amplitude does not work for the daughter trajectories?
A qualitative answer which we could propose is that this amplitude is a
low-energy approach by construction. Successfully describing the
low-energy theorems of current algebra it fails at higher energies
where the physics is different. The spectrum of light mesons in
that region is well described by the effective string theories. The very fact
that the AVW amplitude (unlike the Veneziano one) does not correspond
to any string indicates (although indirectly) that it cannot
describe correctly the high-energy region. This example clearly
shows that direct extrapolations of a low-energy description
(where the pion exchange is of a high importance) to high energies
(where the physics is determined by the gluon exchange which is
chirally invariant) can lead to wrong conclusions about spectral
properties of excited states. For this reason one should be
careful with any statements forbidding CSR which are based on the
language of the low-energy effective field theory (like in~\cite{jaffe}).

The absence of parity doubling on the leading trajectory looks enigmatic.
This fact is rather definite experimentally~\cite{bugg2}. A theoretical
explanation could be the following. The leading $\rho$-meson trajectory
(together with pion) gives the main contribution into the AVW amplitude. 
Therefore, this amplitude
is not very sensitive to the deviations of experimental spectrum for the
daughter trajectories, thus permitting the parity doubling. Hence, the absence
of parity doublets for the states on the leading trajectory seems to be a
minimal price to pay in order to reconcile a good low-energy behavior of
the amplitude with the possibility of parity doubling at higher energies.

\section{Conclusions}

In this Letter we have shown that spectra of baryons and mesons in
the light quark sector have more similarities than one usually
thinks. Namely, like baryons the meson states not only form parity
doublets but also reveal a well pronounced multispin-parity
clustering in the spectrum. These
facts indicate that the phenomena seem to have a unique origin.
The rate of parity doubling in the spectrum of light mesons is
consistent with a genuine chiral symmetry restoration (not only
effective) occurring at $\Lambda_{\text{CSR}}\approx2.5$~GeV. The
scale $\Lambda_{\text{CSR}}$ is a new important scale in QCD which
%should be added to $\Lambda_{\text{QCD}}$ and
supplements $\Lambda_{\text{CSB}}$, i.e the chiral symmetry
breaking phenomenon should be characterized by two scales.
Because of chiral symmetry restoration the Ademollo-Veneziano-Weinberg 
dual amplitude fails in describing the daughter trajectories.
Further experimental search for new light meson resonances is
indispensable for confirmation of above conclusions.

\section*{Acknowledgments}
I would like to thank A. A. Andrianov for critical reading of manuscript
and enlightening comments and also D. V. Bugg for useful correspondence.
The work was supported by CYT FPA, grant 2004-04582-C02-01, CIRIT GC, grant 2001SGR-00065,
RFBR, grant 05-02-17477, and by Ministry of Education and Science of Spain.

\end{document}